\pdfoutput=1
\documentclass[11pt]{article}

\usepackage[final]{acl}

\usepackage{times}
\usepackage{latexsym}

\usepackage[T1]{fontenc}
\usepackage[utf8]{inputenc}

\usepackage{microtype}

\usepackage{inconsolata}

\usepackage{graphicx}

\usepackage{mathtools}
\usepackage[export]{adjustbox}
\usepackage{algorithm2e}
\usepackage{comment}
\usepackage{amsfonts}

\title{Empowering recommender systems using automatically generated Knowledge Graphs and Reinforcement Learning}

\author{
 \textbf{Ghanshyam Verma\textsuperscript{1}},
 \textbf{Simanta Sarkar\textsuperscript{1}},
 \textbf{Devishree Pillai\textsuperscript{1}},
 \textbf{Huan Chen\textsuperscript{1}},
\\
 \textbf{John P. McCrae\textsuperscript{1}},
 \textbf{János A. Perge\textsuperscript{2}},
 \textbf{Shovon Sengupta\textsuperscript{2}},
 \textbf{Paul Buitelaar \textsuperscript{1}},
\\
\\
 \textsuperscript{1}Insight Research Ireland Centre for Data Analytics, Data Science Institute, University of Galway, Ireland \\
 \textsuperscript{2}Fidelity Investments, USA,
\\
 \small{
   \textbf{Correspondence:} \href{mailto:ghanshyam.verma@insight-centre.org}{ghanshyam.verma@insight-centre.org}
 }
}

\begin{document}
\maketitle
\begin{abstract}
Personalized recommendations have a growing importance in direct marketing, which motivates research to enhance customer experiences by knowledge graph (KG) applications. For example, in financial services, companies may benefit from providing relevant financial articles to their customers to cultivate relationships, foster client engagement and promote informed financial decisions. While several approaches center on KG-based recommender systems for improved content, in this study we focus on interpretable KG-based recommender systems for decision-making.

To this end, we present two knowledge graph-based approaches for educational article recommendations for a set of subscribers of a large multinational financial services company. The first approach employs Reinforcement Learning (RL) and the second approach uses the XGBoost algorithm for recommending articles to the subscribers. Both approaches make use of a KG generated from both structured (tabular data) and unstructured data (a large body of text data).

Using the RL-based recommender system we could leverage the graph traversal path leading to the recommendation as a way to generate interpretations (Path Directed Reasoning (PDR)). In the XGBoost-based approach, one can also provide explainable results using post-hoc methods such as SHAP (SHapley Additive exPlanations) and ELI5 (Explain Like I’m Five).

We also compared the above approaches with published SOTA algorithms for building recommender systems. Our proposed RL-based recommender system achieved 43.76\% MAP (MAP@K=10). Our RL-based recommender system outperformed both the XGBoost-based approach and baseline model (Bayesian personalized ranking) by 13.38 and by 32.55 percentage points, respectively, delivering more accurate and personalized educational article recommendations. Importantly, our approach offers explainable results, promoting better decision-making. This study underscores the potential of combining advanced machine learning techniques with KG-driven insights to bolster experience in customer relationship management.
\end{abstract}

\section{Introduction}

The increasing demand for personalized content has led to the development of recommendation systems that can effectively utilize structured information. Knowledge graphs (KGs) have emerged as a promising solution for this challenge, offering improved recommendation performance and explainability due to the inherent comprehensibility of relationships between entities \cite{nickel2011three}. A growing body of research is dedicated to exploring the potential of knowledge graph reasoning in personalized recommendation \cite{a11090137, bordes2013translating, grover2016node2vec, wang2016structural}.
One line of research focuses on knowledge graph embedding models, such as TransE \cite{bordes2013translating} and node2vec \cite{grover2016node2vec}, which align the knowledge graph in a regularized vector space, identifying the similarity between entities by calculating the distance between their representations \cite{zhang2016collaborative}. However, purely KG embedding-based approaches struggle to uncover multi-hop relational paths, limiting the ability to capture complex relationships between entities.
Another line of research investigates path-based recommendation techniques. Gao et al. \cite{gao2018recommendation} proposed the concept of meta-paths for reasoning over KGs. Although promising, this approach faces challenges when dealing with the numerous types of relations and entities present in large, real-world KGs, making it difficult to explore relationships between unconnected entities. Wang et al. \cite{wang2019explainable} developed a path embedding approach for recommendation over KGs that enumerates all qualified paths between every user-item pair, followed by training a sequential RNN model to predict ranking scores for the pairs. While this method improves recommendation performance, it is not feasible to explore all paths for every user-item pair in large-scale KGs due to computational limitations.
Recent advances have focused on combining collaborative filtering (CF) with KG embedding techniques to enhance recommendation performance \cite{zhang2016collaborative, a11090137}. For example, Ai et al. \cite{ a11090137} proposed a method that incorporated a soft matching algorithm to identify explanation paths between users and items. However, this strategy generates explanations post-hoc through empirical similarity matching between user and item embeddings, providing retrospective rationales for the chosen recommendations rather than deriving explanations from the reasoning process \cite{10.1145/3331184.3331203}.
We argue that an intelligent recommendation agent should explicitly reason over knowledge graphs for decision-making rather than simply embedding the graph as latent vectors for similarity matching. In this paper, we treat knowledge graphs as a flexible structure to maintain the agent's knowledge about users, items, other entities, and their relationships. The agent initiates the process with a user and conducts explicit multi-step path reasoning over the graph, discovering suitable items for recommendation. This approach allows for the reasoning process to be easily interpreted, providing causal evidence for the recommended items. Our goal is not only to select a set of candidate items for recommendation but also to provide the corresponding reasoning paths as interpretable evidence for each recommendation.
To address the limitations of previous work, we propose an approach that casts the recommendation problem as a deterministic Markov Decision Process (MDP) over the knowledge graph. We employ a Reinforcement Learning (RL) method, wherein an agent begins with a given user and learns to navigate to potential items of interest. The path in the KG then serves as an explanation for why the item should be recommended to the user. This approach presents three main challenges: measuring the correctness of an item for a user, efficiently exploring promising reasoning paths in the graph, and preserving the diversity of both items and paths during exploration. To tackle these challenges, we propose a KG-driven RL-based approach. The benefit of our approach is that it can also work when reviews or ratings of the items are not available and only click or other forms of response information is available to learn the user preferences.

Our experimental results demonstrate that our proposed method consistently outperforms state-of-the-art recommendation techniques, we present qualitative case studies to demonstrate the explainability of our approach, providing insights into the reasoning paths and decision-making processes of the recommendation agent. These case studies showcase the interpretability of our method, allowing users to better understand the rationale behind the recommendations.
In summary, our research contributes to the growing body of literature on knowledge graph-based recommendation systems, specifically in the financial domain. By proposing a novel reinforcement learning approach and conducting a comparative study with the XGBoost algorithm, we offer valuable insights into the potential of knowledge graphs for improving the performance and explainability of personalized recommendation systems. Our development of a KG-driven XGBoost recommendation system further demonstrates the versatility and applicability of knowledge graph techniques in the field of recommendation.

By developing a KG-driven XGBoost recommendation system alongside our reinforcement learning approach, we aim to showcase the flexibility and potential of knowledge graph-based techniques in addressing a wide range of recommendation scenarios. Our comparative study between the two approaches not only provides insights into their respective strengths and limitations but also highlights the importance of tailoring recommendation algorithms to specific application contexts and requirements. We have made public the source code of both the proposed approaches via a GitHub link\footnote{\url{https://github.com/GhanshyamVerma/Explainable-Recommender-System}.}.

Our main contributions are as follows:
(1) Automatic KG creation using structured and unstructured data. (2) Use of KG for building an XGBoost-based recommender system that can exploit click or other forms of response information. (3) Use of KG for building an RL-based recommender system that can exploit click/response information. (4) Explainability module that can explain the rationale behind the recommendations.

The rest of the paper is structured as follows. In Section~\ref{rel_work}, we describe the existing methods for building recommender systems. Section~\ref{sec_method} describes the methodology. Section~\ref{sec_experimental_setup} describes the experimental setup. In Section~\ref{sec_results}, we discuss and compare results in detail. Finally, we conclude in Section~\ref{sec_conclusion}.

%Through our research, we hope to contribute to the ongoing development and refinement of knowledge graph-based recommendation systems, paving the way for more effective, personalized, and explainable recommendations in various domains. Our work also seeks to inspire further exploration and innovation in the field, fostering the adoption and integration of knowledge graphs and advanced machine learning techniques in the design of next-generation recommendation systems.
\section{RELATED WORK} \label{rel_work}
\subsection{Collaborative Filtering}
Collaborative Filtering (CF) has been a cornerstone in the development of recommender systems. Early approaches to CF focused on the user-item rating matrix and predicted ratings using user-based \cite{konstan1997grouplens, resnick1994grouplens, herlocker1999algorithmic} or item-based \cite{linden2003amazon, sarwar2001item} collaborative filtering methods. These approaches calculated similarities between users or items to generate recommendations.

As dimension reduction methods advanced, latent factor models, such as matrix factorization, gained widespread adoption in recommender systems. Prominent techniques include singular value decomposition \cite{koren2009matrix}, non-negative matrix factorization \cite{lee2000algorithms}, and probabilistic matrix factorization \cite{mnih2007probabilistic}. These methods essentially learn a latent factor representation for each user and item to calculate the matching score of user-item pairs.

In recent years, deep learning and neural models have further extended collaborative filtering, leading to two main sub-categories: similarity learning and representation learning. The similarity learning approach adopts relatively simple user/item embeddings (e.g., one-hot vectors) and learns a complex prediction network as a similarity function to compute user-item matching scores \cite{he2017neural}. In contrast, the representation learning approach focuses on learning richer user/item representations, while using a simple similarity function (e.g., inner product) for score matching \cite{zhang2017joint, wang2019neural}.

However, the recommendation results generated by latent factor or latent representation models can be difficult to explain, which has led to a growing interest in explainable recommendation [19, 20]. The challenge of making recommendations more interpretable has driven researchers to explore various techniques and approaches that offer both high-quality recommendations and meaningful explanations for the user-item associations.

In response to the challenges posed by the lack of interpretability in traditional collaborative filtering approaches, researchers have started to explore hybrid recommender systems that combine the benefits of CF methods with other techniques, such as knowledge graph-based methods \cite{guo2020survey, zhang2016collaborative}. These hybrid systems aim to improve the quality of recommendations while also providing more interpretable and explainable results.
 
Knowledge graphs provide a structured representation of information, making it easier to reason about the relationships between entities and draw meaningful connections. By incorporating knowledge graphs into the recommendation process, researchers can develop systems that offer both high-quality recommendations and interpretable explanations for user-item associations.

The field of collaborative filtering-based recommender systems has seen significant advancements over the years, with a growing emphasis on integrating additional sources of information and enhancing interpretability. The exploration of hybrid systems, such as those that combine collaborative filtering with content-based filtering or knowledge graph-based methods, holds promise for the development of more accurate, personalized, and explainable recommendations.

\subsection{Knowledge Graph-driven Recommender Systems}
Knowledge Graph-driven Recommender Systems (KGRS) have recently gained attention due to their ability to provide explainable and high-quality recommendations. Researchers have explored different ways to incorporate knowledge graph embeddings into recommender systems to improve recommendation performance and interpretability.
One research direction focuses on leveraging knowledge graph embeddings as rich content information to enhance recommendation performance. For example, Zhang et al. \cite{zhang2016collaborative} utilized knowledge base embeddings to generate user and item representations for recommendation purposes. Huang et al. \cite{huang2018improving} employed memory networks over knowledge graph entity embeddings for recommendation. Wang et al. \cite{wang2018ripplenet} proposed a ripple network approach for embedding-guided multi-hop KG-based recommendation, which allows for the exploration of connections between entities in the knowledge graph.
Another research direction aims to leverage the entity and path information in the knowledge graph to make explainable decisions. Ai et al. \cite{a11090137} incorporated the learning of knowledge graph embeddings for explainable recommendation, but their explanation paths are essentially post-hoc explanations, as they are generated by soft matching after the corresponding items have been chosen. Wang et al. \cite{wang2019explainable} proposed an RNN-based model to reason over KGs for recommendation. However, this approach requires enumerating all possible paths between each user-item pair for model training and prediction, which can be impractical for large-scale knowledge graphs.

The field of Knowledge Graph-driven Recommender Systems has witnessed significant progress in recent years. Researchers are exploring different approaches to incorporate knowledge graph embeddings and entity relationships to enhance recommendation performance while providing interpretable and explainable results. Future work in this area will likely focus on developing more efficient and scalable methods for reasoning over large-scale knowledge graphs and further improving the quality and explainability of recommendations.

Some researchers have focused on leveraging the structural properties of knowledge graphs to improve recommendation performance. For instance, Wang et al. \cite{wang2019neural} developed a graph attention network that incorporates both the relational information and entity features in a knowledge graph for recommendation. This approach allows for more accurate and context-aware recommendations by attending to the most relevant relations and entities for a given user-item pair.

In addition to using knowledge graph embeddings, researchers have also explored incorporating external knowledge sources and incorporating user-item interactions into the knowledge graph. Cao et al. \cite{cao2019unifying} proposed a unified framework for incorporating user-item interactions and external knowledge sources into the knowledge graph, which improved the quality of recommendations by capturing the complex interplay between these elements.

Schlichtkrull et al. \cite{schlichtkrull2018modeling} introduced a relational graph convolutional network (R-GCN) that learns embeddings for both entities and relations in a knowledge graph. This method can be used in a wide range of applications, including recommender systems, by exploiting the rich information present in the knowledge graph structure.

The research area of Knowledge Graph-driven Recommender Systems has experienced significant advancements, with researchers exploring various methods to utilize knowledge graph embeddings, external knowledge sources, and user-item interactions to improve the quality and explainability of recommendations. As more efficient and scalable techniques are developed, KGRS will continue to evolve and provide increasingly accurate, personalized, and explainable recommendations.

\subsection{Reinforcement Learning based Recommender Systems}
Reinforcement Learning (RL) has garnered considerable interest in the research community, with numerous successful applications in various domains, including recommender systems. Researchers have explored RL-based recommender systems in both non-KG settings and KG settings for a range of tasks.

In non-KG settings, RL has been applied to various types of recommender systems, such as ads recommendation \cite{theocharous2015ad}, news recommendation \cite{zheng2018drn}, and post-hoc explainable recommendation \cite{wang2018reinforcement}. These applications have demonstrated the potential of RL to adapt to changing user preferences and generate personalized recommendations based on user interactions.

In the context of knowledge graphs, researchers have primarily focused on utilizing RL for tasks such as question-answering (QA). For instance, Xiong et al. \cite{xiong2017deeppath} leveraged reinforcement learning for path-finding in knowledge graphs, while Das et al. \cite{das2017go} proposed MINERVA which makes use of a KG and trains a model for question answering. Lin et al. \cite{DBLP:conf/emnlp/LinSX18} introduced RL-based models for KG question answering with reward shaping. These approaches formulate multi-hop reasoning as a sequential decision-making problem, taking advantage of the structure and information present in knowledge graphs.

However, to the best of our knowledge, there has been limited research on utilizing RL in knowledge graphs specifically for the task of recommendation, especially when considering the challenge of navigating an extremely large action space as the number of path hops grows. This opens up a promising research direction for developing RL-based recommender systems that can exploit the rich information present in knowledge graphs while efficiently navigating large action spaces to provide personalized and explainable recommendations.

Reinforcement learning presents a promising avenue for recommender systems, particularly when combined with the rich information present in knowledge graphs. By exploring novel techniques for managing large action spaces, incorporating graph neural networks, and leveraging transfer learning, researchers can continue to push the boundaries of RL-based recommender systems, providing increasingly accurate, personalized, and explainable recommendations.

\section{METHODOLOGY} \label{sec_method}
The problem addressed in this research is to provide a new type of recommendation, called Knowledge Graph Driven Explainable Recommendation (KGDExR), that simultaneously performs item recommendation and path finding based on rich and heterogeneous information in the knowledge graph. 

The goal is to find a recommendation set of $N$ items for a given user $u$ from a subset of Item entities $\mathbf{I}$ connected to User entities $\mathbf{U}$ through relations $r_{ui}$ in The knowledge graph $\mathbf{G}$. The recommendation set should be associated with one reasoning path $p_{j}(u,i_{n})$ $(2 \leq j \leq J)$ for each pair $(u,i_{n})$ of user and recommended item, where $j$ is the number of hops in the path and $J$ is a given integer. The number of recommendations, $N$, is also given as an input.
The knowledge graph $\mathbf{G}$ is defined as $\mathbf{G} = (e^{h},r, e^{t})$, where $e^{h}$ is the head entity and $e^{t}$ is the tail entity in the KG. $e^{h}$ \& $e^{t}$ $\in \mathbf{E}, r \in \mathbf{R}$, where $\mathbf{E} $ is the entity set and $\mathbf{R}$ is the relation set. A j-hop path from entity $e_{0}$ to entity $e_{j}$ is defined as a sequence of $j+1$ entities connected by $j$ relations, denoted by $ p_{j}(e_{0},e_{j}) = \left \{ e_{0} \overset{r_{1}}{\leftrightarrow }   e_{1} \overset{r_{2}}{\leftrightarrow } . . . \overset{r_{j}}{\leftrightarrow } e_{j} \right \} $.

The KGDExR problem can be formalized as finding a set of $N$ items $\left \{i_{n}\right \}_{n \in [\mathbf{N}]} \subseteq  \mathbf{I}$ for a given user $u$ and integers $J$ and $N$, such that each pair $(u,i_{n})$ is associated with a reasoning path $p_{j}(u,i_{n})$ $(2 \leq j \leq J)$. 

\subsection{KG-Driven Reinforcement Learning based Recommender System}

We use Markov Decision Process (MDP) framework to address the KGDExR problem. To ensure path connectivity, we supplement the graph $\mathbf{G}$ with two distinct types of edges. Primarily, reverse edges are included, such that if $(e^{h},r, e^{t}) \in \mathbf{G}$, then $(e^{t},r, e^{h}) \in \mathbf{G}$, aiding in the path definition.

The state at a given step $t$, denoted as $s_{t}$, is represented as a triplet $(e_{u}, e_{s_{t}}, h_{t})$, where $e_{u} \in U$ denotes the initial user entity, $e_{s_{t}}$ indicates the entity the agent has arrived at step $t$, and $h_{t}$ refers to the history before step $t$. We define the k-step history as the combination of all entities and relations in the previous k steps, i.e., $\left \{ e_{u} \overset{r_{j}}{\leftrightarrow }   e_{j} \overset{r_{j+1}}{\leftrightarrow } . . . \overset{r_{j+k-1}}{\leftrightarrow } e_{k-1} \overset{r_{j+k}}{\leftrightarrow } e_{k} \right \} $. Given some user $u$, the initial state is represented as $s_0 = (e_{u}, e_{u}, \emptyset)$ and the terminal state is represented as  $s_{T} = (e_{u}, e_{T}, h_{T})$.

The action space $A_{t}$ at state $s_{t}$ is defined as all possible emerging edges from an entity $e{t}$.
Some nodes in the KG can have very large out-degree which can make it inefficient to maintain the large action space. Therefore, we perform an action-pruning step based on a scoring function $f((r,e)|u)$, which maps any relation to a real-valued score conditioned on a given user \cite{xian2019reinforcement}. There is a user-defined integer $\alpha$ that upper bounds the size of the action space. For our experiments, we set the value of $\alpha$ = 3.

For a given user, a simple binary reward function is not appropriate as we don't know whether the agent has reached a target item or not. Therefore, the agent needs to find as many reasoning paths as possible. We consider giving a reward to the last state ($s_{T}$) of the path. The reward $R_{T}$ is defined as:
\begin{equation}
R_{T} = \begin{cases*}
  max\left ( 0, \frac{f(u,e_{T})}{max_{i\in I} f(u,i)} \right ), & if $ e_{T}\in I$,\\
  0,                    & otherwise.
\end{cases*}
\end{equation}

In accordance with the underlying properties of the graph, the state in our recommendation system is determined by the entity's position. Given a state $s_{t} = (e_{u}, e_{t}, h_{t})$  and an action $a_{t} = (r_{t+1}, e_{t+1})$, the transition to the next state $s_{t+1}$ is characterized by a probability distribution:

\begin{equation}
\begin{split}
P[s_{t+1} = (e_{u}, e_{t+1}, h_{t+1}) | s_{t} = \\
(e_{u}, e_{t}, h_{t}), a_{t} = (r_{t+1}, e{t+1})] = 1
\end{split}
\end{equation}

However, there is an exceptional case for the initial state $s_{0} = (e_{u}, e_{u}, 0)$, which introduces stochasticity and depends on the starting user entity. To simplify the model, we assume a uniform distribution for the users, ensuring that each user is equally sampled at the beginning.

Building upon our Markov Decision Process (MDP) formulation, our primary objective is to learn a stochastic policy $\pi$ that maximizes the expected cumulative reward.

We define the expected cumulative rewards based on all the paths traversed by a user as below:

\begin{equation}
J(\theta ) = \mathbb{E}_{e_{0}}\in u [\mathbb{E}_{a_{1}, a_{2}, ..., a_{T} \sim \pi_{\theta}(a_{t}|s_{t})} [R_{T}]]
\end{equation}

To maximize the expected cumulative rewards, we use gradient ascent. The gradients are derived by the REINFORCE \cite{sutton2018reinforcement}, i.e.,

\begin{equation}
\bigtriangledown_{\theta} J(\theta) \approx \bigtriangledown_{\theta} \sum_{t} R_{T}  log  \pi_{\theta}(a_{t}|s_{t}).
\end{equation}

The final step of our recommendation problem solution involves using a trained policy network to guide the exploration of a knowledge graph. Our objective is to find a set of candidate items and their corresponding reasoning paths for a given user. One approach is to sample paths for each user based on the policy network's guidance. However, this method may lack path diversity because the agent tends to repeatedly search the path with the highest cumulative rewards. To address this, we propose using Path Directed Reasoning (PDR) algorithm, which considers both action probability and reward, to explore candidate paths and recommended items for each user. The process is outlined in Algorithm 1. The algorithm takes inputs such as the KG, the user, and the policy network. The output is a set of T-hop paths for the user, along with their generative probabilities and rewards. Each path ends with an item entity and associated generative probability and reward. Among the candidate paths, there may be multiple paths between the user and an item. To interpret the reasoning behind why an item is recommended to the user, we select the path from the candidate set with the highest generative probability based on the generative probabilities. Finally, we rank the selected interpretable paths based on their path rewards and recommend the corresponding items to the user.

\RestyleAlgo{ruled}

%% This is needed if you want to add comments in
%% your algorithm with \Comment
\SetKwComment{Comment}{/* }{ */}

\begin{algorithm}[hbt!]
\caption{Path Directed Reasoning (PDR) Algorithm }\label{alg:PDR}
\KwData{KG $G$, items $I$, users $U$; policy $\pi$}
\KwResult{Reward $R$ ; path $P$; probability $X$ }
Initialize $R$, $P$ and $X$\;
\For{ all $u \in U$}{
    \For{$t = 1$ to $T$}{
    Initialize $R_{t} = \phi$, $P_{t} = \phi$, $X_{t} = \phi$ \;
        \For{$\hat{p} \in P$, $\hat{r} \in R$, $\hat{x} \in X$}{
        Path $\hat{p} = \left\{e_{u}, r_{1}, ... , r_{t-1}, e_{t-1}\right\}$\;
        Set state $s_{t-1} = (e_{u}, e_{t-1}, h_{t-1})$\;
        Get pruned action space $\hat{A}_{t-1}(u)$\;
        Get a path for action $a$ such that $p(a) = \pi(a|s_{t-1})$\;
        Actions $A_{t} = \left\{a| rank(p(a))\right\}$\;
            \For{all $a \in A_{t}$}{
            Get state $s_{t}$ and $R_{t}$\;
            Assign new path  $\hat{p} \cup \left\{r_{t}, e_{t}\right\}$ to $P_{t}$\;
            Assign new probability  ${p(a)\hat{x}}$ to $X_{t}$\;
            Assign new reward  $ R_{t+1}+\hat{r}$ to $R_{t}$\;
            }
    
        }

    }
    Export all paths that end with an item $i \in I$\;
    Return updated $P_{T}, X_{T}$ and $R_{T}$ \;
}
\end{algorithm}

\subsection{KG-Driven XGBoost based Recommender System}
XGBoost (eXtreme Gradient Boosting)~\cite{10.1145/2939672.2939785} is an ensemble learning algorithm that has become a popular and effective method for a wide range of machine learning tasks, including classification, regression, and ranking. XGBoost builds a set of decision trees iteratively, using a gradient boosting approach to minimize a user-specified loss function.

For a dataset $D = \left\{ (\textbf{x}_{i}, y_{i}) \right\} |  
 (\textbf{x}_{i} \in \mathbb{R}^m, y_{i} \in \mathbb{R} )$ that has $n$ observations and $m$ features, the XGBoost model uses $Z$ additive functions for prediction~\cite{10.1145/2939672.2939785}.
\begin{equation}
\hat{y}_{i} = \sum_{z=1}^{Z} f_{k} (\textbf{x}_{i}),
\end{equation}
where $f_k \in F$ and $F$ is the space of regression trees which can be defined as:
\begin{equation}
F = \left\{f(\textbf{x}) = w_{q(x)} \right\} (q: \mathbb{R}^m \to T, w \in \mathbb{R}^T ),
\end{equation}
where $q$ is the structure of each tree that maps an observation to the corresponding leaf node in the tree, $T$ represents the number of leaf nodes in the tree, and $w$ represents the leaf weights. For a given observation, the final prediction is computed by taking the sum of all the weights for the corresponding leave nodes.

The key idea behind XGBoost is to iteratively add decision trees to the ensemble, with each new tree trained to correct the residual errors of the previous trees. In other words, XGBoost fits the model by adding new trees to the ensemble that improve the overall prediction accuracy, while penalizing trees that are too complex or overfit the data.

One of the important features of XGBoost is its support for a wide range of objective functions and evaluation metrics, including common loss functions like squared error and logistic loss, as well as custom loss functions. XGBoost also includes a variety of regularization techniques to prevent overfitting and improve generalization performance, including L1 and L2 regularization terms, tree depth constraints, and early stopping.

For our initial experiments, we implemented three rankers within the XGBoost model to predict the ranking of the articles for the users. These are XGBoost ranker~\cite{10.1145/2939672.2939785}, CatBoost ranker \cite{prokhorenkova2018catboost}, and LightGBM ranker \cite{ke2017lightgbm}. CatBoost \cite{prokhorenkova2018catboost} is a recent library known for its efficacy in handling categorical features, which employs YetiRank \cite{gulin2011winning} as the loss function. LightGBM \cite{ke2017lightgbm} handles categorical features and optimizes the LambdaRank loss. We trained XGBoost ranker ~\cite{10.1145/2939672.2939785} with Rank Pairwise loss, utilizing one-hot encoding. During our initial experiments, the XGBoost ranker outperformed the other two rankers. Therefore, we selected the XGBoost ranker for our KG-driven XGBoost-based recommender system approach.

We used XGBoost ranker in combination with KGs generated from article text and the other article features to build the XGBoost-based recommender system. The KGs generated are then used as input to the TuckER and TransE to generate 300-dimensional KG embeddings. These embeddings along with the subscriber demographical data and educational article features are used to train the KG-driven XGBoost-based recommender system.  

\section{Experimental Setup} \label{sec_experimental_setup}

In this section, we provide information on KG creation, KG embedding generation, and the data sets used in this work.

\subsection{Automatic KG Generation}
To automatically generate KGs from the targeted unstructured data sets, we used two approaches. The first approach makes use of external lexical resources, such as ConceptNet \cite{speer2017conceptnet} to connect terms and enrich the taxonomy. The second approach is different in the way that it neither requires any training nor any external resource, but instead uses the knowledge of the domain available within the input data to extract the knowledge.

\subsubsection{ConceptNet-based approach} ConceptNet \cite{speer2017conceptnet} is a knowledge graph that encompasses entities from various domains along with their corresponding relationships. For this study, we specifically focus on three relationship types: IsA, PartOf, and Synonym. The "IsA" relationship signifies hypernymy relations, while "PartOf" represents meronymy relations, and "Synonym" indicates synonymy relations. To generate a dataset for hyponymy relations, we inverted the direction of relations labeled as hypernyms. All other relations in ConceptNet were grouped together as "other." The training dataset was created by including all extracted relationships.

The system architecture is based on BERT \cite{DBLP:conf/naacl/DevlinCLT19}, employing 12 transformer blocks. The embeddings utilized are extracted from the transformer in the 12th layer. Pretrained embeddings from the BERT model "uncased\_L-24\_H-1024\_A-16" are employed, which are readily available in TensorFlow. We named "uKG\_CN" to the KG that we generated using the ConceptNet-based approach.

\subsubsection{Dependency Parsing-based approach}
The creation of a domain-specific KG with this approach follows a mixed approach based on both the Saffron tool\footnote{\url{https://saffron.insight-centre.org/}} for taxonomy generation and the new algorithm for relation extraction. It uses the syntactic knowledge of sentences in a textual dataset to extract new relations between Saffron terms. After extracting the new relations from the text, we integrate them into the Saffron taxonomy and return a fully formed KG. This approach does not require any training and is domain independent.

The dependency parsing-based relation extraction approach extracts relations from the text and exports them as triples (left\_relation, relation\_type, right\_relation). It uses dependency parsing (syntactic analysis of the sentences) on the text to find how terms are syntactically (and by extension semantically) connected within sentences. It takes as input the terms extracted by Saffron \citep{Pereira2019}, as well as the dataset originally used to extract the Saffron terms and extract the taxonomy, and returns a list of triples: {term1, relation, term2}. The whole implementation is done in Python. We named "uKG\_DP" to the KG that we generated using the Dependency Parsing based approach.

We have also created a KG, referred to as "uKG", from unstructured data. This KG contains only the article and its relation with the most frequent terms found within the article. To compute the Term Frequency, we utilized TF-IDF.

\subsubsection{KG creation using both structured and unstructured data (cKG)}
We have already defined the (KGDExR) problem and provided the definition of a KG in section \ref{sec_method}. Here, we will illustrate how we constructed KG using both structured and unstructured data (combined data (cKG)). The features of structured data, such as `subscriber', `educational\_article', `topic', `product', `topic\_tag', `product\_tag', `response', etc., serve as the type of nodes or entities in the KG. These entities are connected to other entities through relations such as `has\_topic', `has\_product', `has\_topic\_tag', `has\_product\_tag', and `has\_response'. Additionally, we utilized the full text of the article, which represents the unstructured data, to create this KG. Therefore, this KG leverages both structured and unstructured data for its creation. The recommendation process begins with a subscriber, traverses through specific entities and their associated relations, and ultimately leads to an item, which in our case is the recommended educational article for that subscriber. We have named the KG generated using structured and unstructured data that is the combined data as "cKG".

\subsection{Knowledge Graph Embeddings}

In a given KG, each head entity or tail entity can be associated as a point in a continuous vector space. In this work, we use TuckER \citep{balazevic2019tucker, info14050288} and TransE \cite{wang2014knowledge} methods to generate KG embeddings. TuckER employs a three-way TuckER tensor decomposition, which computes the tensor T and a sequence of three matrices leveraging the embeddings of entities ($E_{head}$ and $E_{tail}$) and relations ($R$) between them ($G \approx T \otimes E_{head} \otimes R \otimes E_{tail}$). 

The underlying idea of TransE is to interpret relations as translations that occur between entities in the knowledge graph. In TransE, each entity and relation is assigned a unique vector representation in the embedding space. The objective of the model is to learn these embeddings in such a way that the translation between the embeddings of a head entity and a relation should be close to the embedding of a tail entity. These methods allow us to create KG embeddings that are used to train our recommender systems.

\subsection{Data sets}

The dataset used in this study contains the data of the subscribers of a large multinational financial services company and the educational articles sent to these subscribers. The dataset spans from January 30th, 2019 to October 30th, 2019, and contains information of 463 subscribers who opted for approximately 80 articles each during this period.
The dataset consists of 37,423 rows, detailing individual subscriber-article interactions. It includes a total of 71 educational articles, with 66 unique articles, providing details related to financial products and services. This dataset serves as a valuable resource for researchers and marketers interested in understanding subscriber's behavior and preferences and choices made by them, as well as identifying opportunities for targeted content and marketing strategies. We used this dataset for the evaluation of our KG-driven RL-based approach and KG-driven XGBoost approach for recommending educational articles to subscribers. The dataset is divided into training, and test sets with a ratio of 70:30 respectively. We have also made this data set publicly available on a GitHub repository~\footnote{\url{https://github.com/GhanshyamVerma/Explainable-Recommender-System}.}.

\begin{table*}
\caption{Results of KG-driven XGBoost based Recommender system and KG-driven RL based Recommender system with baseline XGBoost approach. }
\begin{adjustbox}{width=.99999\textwidth,totalheight=\textheight,keepaspectratio}
\label{table_1_embedding}
\begin{tabular}{|l|l|c|c|c|}
\hline
\textbf{Model}                                                                          & \textbf{Embedding}                                                                              & \multicolumn{1}{l|}{\textbf{MAP@K=10}} & \multicolumn{1}{l|}{\textbf{Precision@K=10}} & \multicolumn{1}{l|}{\textbf{Recall@K=10}} \\ \hline
\begin{tabular}[c]{@{}l@{}}BPR \\ (Bayesian personalized ranking)\end{tabular} & \multicolumn{1}{c|}{-}                                                                          & 0.11207                              & 0.05672                                 & 0.41904                              \\ \hline
\begin{tabular}[c]{@{}l@{}}Neighborhood-based \\ Recommender System\end{tabular} & \multicolumn{1}{c|}{-}                                                                          & 0.20418                              & 0.66177                                 & 0.27175
                              \\ \hline
\begin{tabular}[c]{@{}l@{}}NCF  \\ (Neural Collaborative Filtering)\end{tabular} & \multicolumn{1}{c|}{-}                                                                          & 0.24104                              & 0.62599                                 & 0.30007                              \\ \hline
XGBoost                                                                        & \begin{tabular}[c]{@{}l@{}}Sentence Transformer Embedding\\ {[}all-MiniLM-L6-v2{]}\end{tabular} & 0.30381                              & 0.65902                                 & 0.21568                             \\ \hline
\begin{tabular}[c]{@{}l@{}}KG-XGBoost \\ {[} uKG\_DP {]}\end{tabular}          & \begin{tabular}[c]{@{}l@{}}Saffron Dependency Parsing KG \\ (TuckER KG Embedding)\end{tabular}  & 0.34378
                              & 0.71708                                 & 0.23965                             \\ \hline
\begin{tabular}[c]{@{}l@{}}KG-XGBoost \\ {[} uKG\_CN {]}\end{tabular}          & \begin{tabular}[c]{@{}l@{}}Saffron ConceptNet KG \\ (TuckER KG Embedding)\end{tabular}         & 0.38985
                              & 0.24575                                 & 0.74384                              \\ \hline
\begin{tabular}[c]{@{}l@{}}KG-XGBoost \\ {[} uKG {]}\end{tabular}              & TransE KG Embedding                                                                             & 0.33774                              & 0.23137                                 & 0.70987                              \\ \hline
\begin{tabular}[c]{@{}l@{}}KG-XGBoost \\ {[} cKG {]}\end{tabular}                   & TransE KG Embedding                                                                             & 0.34468
                              & 0.24031                                 & 0.73740                             \\ \hline
\textbf{\begin{tabular}[c]{@{}l@{}}KG-RL \\ {[} cKG {]}\end{tabular}}                   & TransE KG Embedding                                                                             & 0.43761                              & 0.60562                                 & 0.24857                              \\ \hline
\end{tabular}
\end{adjustbox}
\end{table*}

% \begin{table*}[]
% \caption{Results of KG-RL and KG-XGBoost based recommender systems with baseline approaches. }
% \label{table_2_compare_baseline}
% \begin{tabular}{|l|ccc|}
% \hline
% \textbf{Dataset}                                                               & \multicolumn{3}{c|}{\textbf{Customer-Article Data}}            \\ \hline
% \textbf{Measures (\%)}                                                         & \multicolumn{1}{c|}{\textbf{MAP@10}} & \multicolumn{1}{c|}{\textbf{Precision}} & \textbf{Recall}  \\ \hline
% \textbf{BPR (Bayesian personalized ranking)}                                   & \multicolumn{1}{c|}{0.11207}        & \multicolumn{1}{c|}{0.05672}        &   0.41904       \\ \hline
% \textbf{XGBoost}                                                               & \multicolumn{1}{c|}{0.32806} & \multicolumn{1}{c|}{0.20685} & 0.63527 \\ \hline
% \textbf{KG-XGBoost}                                                            & \multicolumn{1}{c|}{0.33047} & \multicolumn{1}{c|}{0.23097} & 0.69039  \\ \hline
% \textbf{KG-RL}                                                                 & \multicolumn{1}{c|}{0.43761} & \multicolumn{1}{c|}{0.60562} & 0.24857  \\ \hline
% \end{tabular}

% \end{table*}

\section{Results} \label{sec_results}

We have produced results using both KG-driven XGboost approach and KG-driven reinforcement learning approach.

Table \ref{table_1_embedding} represents the results obtained using the proposed approaches with the KG embeddings used for the model building. From Table \ref{table_1_embedding}, we can see that the baseline XGBoost model with sentence transformer embedding [all-MiniLM-L6-v2] achieved a 30.38\% MAP score. We observed improvements in performance when we used KG embeddings compared to when KG embeddings were not used (see Tabel \ref{table_1_embedding}). 

We constructed two KGs using unstructured data (educational article contents/texts) through Saffron \citep{Pereira2019} as mentioned in Section \ref{sec_experimental_setup}. These KGs are "uKG\_DP" and "uKG\_CN" where u denotes unstructured data, DP denotes dependency parsing and CN denotes ConceptNet. Additionally, we created a KG referred to as "cKG" from both structured and unstructured data, as explained in Section \ref{sec_experimental_setup}. 

The rationale behind using the cKG with RL-based approach is that it helps in generating explainable recommendations using paths in the cKG. For RL based approach we used KG embeddings generated using TransE, as shown in Table \ref{table_1_embedding}.

We also compared the performance of our proposed approaches with state-of-the-art (SOTA) existing recommender systems. The existing recommender systems we used are: BPR (Bayesian personalized ranking), Neighborhood-based Recommender System, NCF
(Neural Collaborative Filtering), and XGBoost with sentence embedding. We observed that BPR achieved a MAP score of 11.21\%, whereas the KG-driven XGBoost approach (cKG) and KG-driven RL-based approach using the same cKG achieved 34.47\% and 43.76\% MAP scores, respectively. The KG-driven XGBoost approach with KG generated using ConceptNet achieved a MAP score of 38.98\% with a recall of 74.38\%. The results suggest that if recall is important for any application, then KG-driven XGBoost with uKG\_CN can be considered as an option, as it provides the highest recall. Based on the results, it can be observed that the KG-driven RL-based approach outperformed the BPR, Neighborhood-based
Recommender System, NCF, and KG-driven XGBoost approaches when considering the MAP score. Additionally, among all the experiments conducted with KG embeddings, the KG embeddings generated from TransE have proven to capture useful information, resulting in better performance compared to TuckER embeddings.

\begin{figure}[h]
  \centering
  \includegraphics[width=\linewidth]
  {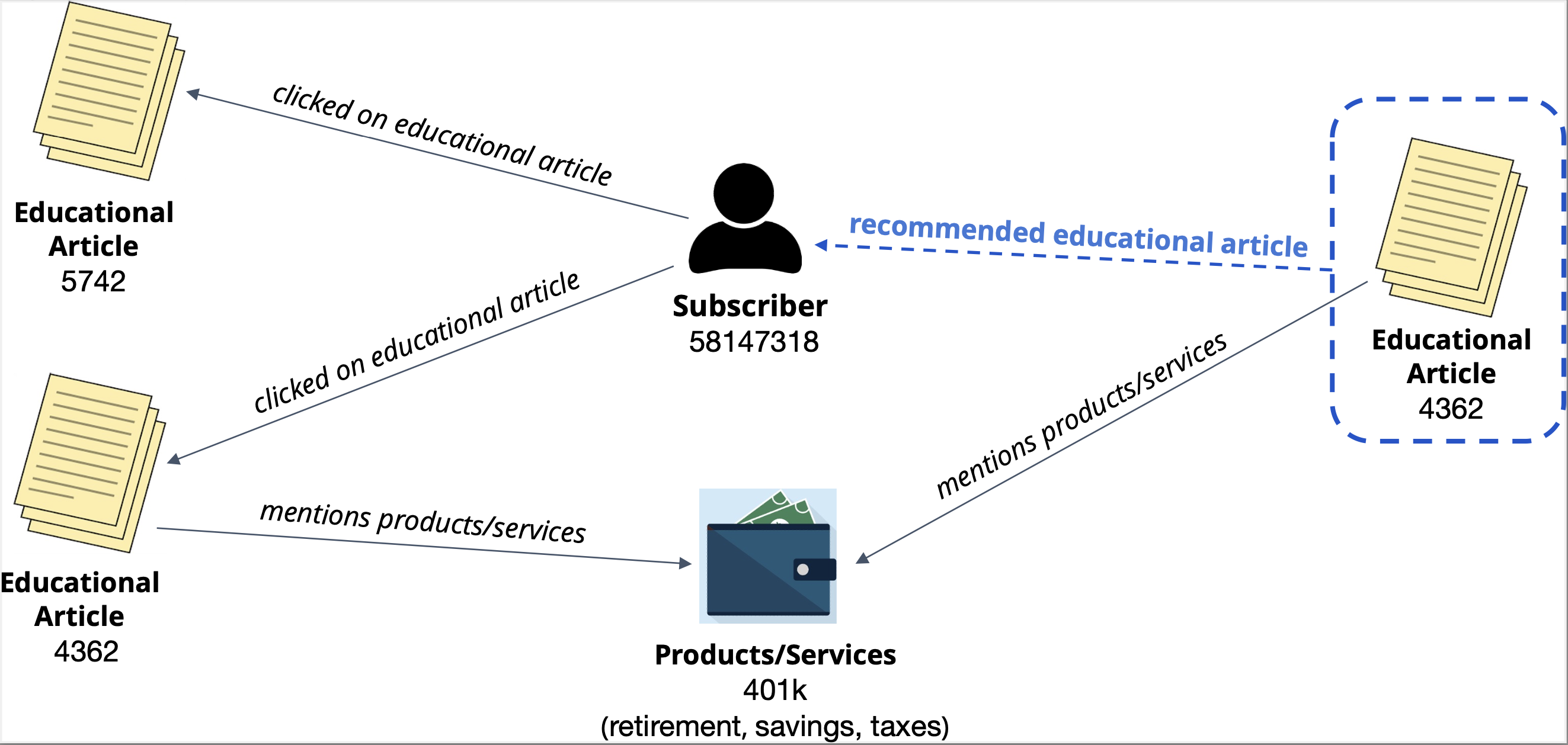}
  % {figures/Figure_explain_results.pdf}
  \caption{Explaining the recommendations of RL-based approach using the path in the KG that leads to the recommendation.}
  \label{fig1}
\vspace{-0.5em}
\end{figure}

% \begin{figure}
% \centering
% \vspace{-7em}
% \includegraphics[width=0.54\textwidth]{figures/xgboost_kg_cn100_shap.pdf}
% \vspace{-7em}
% \caption{Explaining the recommendations of XGBoost-based approach using SHAP.}
% \label{fig_xgboost_kg_cn100_shap}
% \end{figure}

\begin{figure}
%\vspace{-12em}
\adjustimage{width=.45\textwidth}
{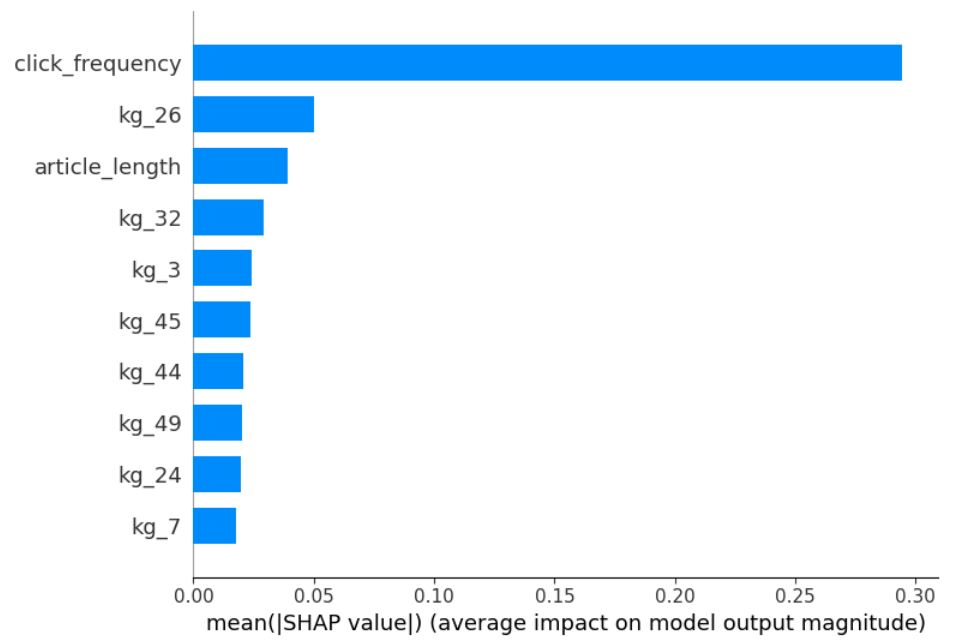}
% {figures/xgboost_kg_cn100_shap_v2.JPG}
%\vspace{-18em}
\caption{Explaining the recommendations of KG-XGBoost [uKG\_CN] model using SHAP.}
\label{fig_xgboost_kg_cn100_shap}
\end{figure}

% \begin{figure}
% \vspace{-12em}
% %\includegraphics[width=0.70\textwidth, left]
% {figures/xgboost_kg_cn100_eli5.pdf}
% \adjustimage{width=.7\textwidth,left}{figures/xgboost_kg_cn100_eli5.pdf}
% \vspace{-18em}
% \caption{Explaining the recommendations of XGBoost-based approach using ELI5.}
% \label{fig_xgboost_kg_cn100_eli5}
% \end{figure}

Our KG-driven RL-based approach is explainable. To gain a better understanding of our model's interpretation of the recommendation, we present a case study based on the results obtained from our experiments. In this study, we analyze the path patterns uncovered by our model during the reasoning process, as well as examine different recommendation scenarios. As shown in Figure \ref{fig1}, the educational article highlighted with a blue dashed boundary is the article recommended by our RL-based model to a subscriber. We can see that the recommended article has some similarities with another educational article already recommended and clicked by that subscriber, therefore the model thinks that this article should be of relevance for that subscriber as the subscriber was interested in such kind of articles before. Furthermore, our RL-based approach enables us to offer the top 10 educational articles for each subscriber. Additionally, it can provide all the associated articles in the path that lead to the outcome, along with shared products/services, topics, and the most frequent common terms found in the text of the educational articles present in the path. Our RL-based approach can provide such paths for each recommended item to a user which explain the results and play an important role in decision-making. 

\begin{figure}
%\vspace{-12em}
\adjustimage{width=.25\textwidth}
% {figures/xgboost_kg_cn100_eli5_v2.JPG}
{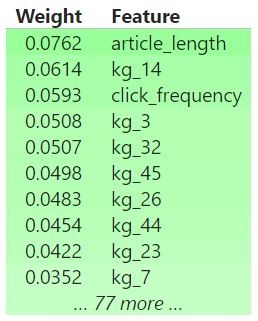}
%\vspace{-18em}
\caption{Explaining the recommendations of KG-XGBoost [uKG\_CN] model using ELI5.}
\label{fig_xgboost_kg_cn100_eli5}
\end{figure}

To generate post-hoc explanation for KG driven XGBoost-based approach, we used SHAP \cite{NIPS2017_7062} and ELI5\footnote{\url{https://github.com/TeamHG-Memex/eli5}}. SHAP (SHapley Additive exPlanations) is a model-agnostic method used for explaining the output of machine learning models. It is based on game theoretic concepts and provides an explanation for each feature's contribution to the model's prediction. SHAP values quantify the impact of each feature by assigning a value to it, indicating how much it contributes to the prediction compared to the average prediction. SHAP relies on the concept of Shapley values from cooperative game theory and it considers additive feature importance.
% {These values are derived from the "Shapley values" - - a concept in cooperative game theory.} 
Figure \ref{fig_xgboost_kg_cn100_shap} represents the KG-XGBoost [uKG\_CN] model's features with their average impact on the model output generated by SHAP. 

ELI5 (Explain Like I'm 5) is a Python library or framework for explainable machine learning models. ELI5 focuses on understanding the overall behavior and importance of features in making predictions. Eli5 reports feature importance using the "permutation importance" algorithm. Figure \ref{fig_xgboost_kg_cn100_eli5} shows the KG-XGBoost [uKG\_CN] model's feature importance by assigning weights to the features based on their impact on the model output generated by ELI5. Both SHAP and ELI5 show that click\_frequency, kg\_26, article\_length, kg\_32, Kg\_3, and Kg\_45 are the most important features that contributed most to the model results.

Overall, the proposed approaches are helpful in providing insights to understand the recommendations and simultaneously perform better than the existing baseline recommender systems.

\section{Conclusion} \label{sec_conclusion}
This research paper explores and demonstrates the use of knowledge graphs (KGs) to enhance personalized recommendations in the financial sector. We proposed two KG-driven recommender systems for a large multinational financial services company, utilizing reinforcement learning and the XGBoost algorithm, respectively. The first approach employs Reinforcement Learning (RL), while the second utilizes the XGBoost algorithm. The XGBoost-based approach uses KG embeddings generated from both TuckER and TransE, and the RL-based approach uses TransE-generated embeddings. We also performed experiments keeping the KG and the embedding same. The findings suggest that the KG-driven RL-based approach outperforms both the KG-driven XGBoost system and baseline models, delivering more accurate and personalized educational article recommendations. Additionally, the study emphasizes the importance of reasoning with knowledge for decision-making. Overall, this study highlights the potential of combining advanced machine learning techniques with KG-driven insights to improve customer experience and drive business growth in the investment sector.

% \subsection{Appendices}

% Use \verb|\appendix| before any appendix section to switch the section numbering over to letters. See Appendix~\ref{sec:appendix} for an example.

% \section{Bib\TeX{} Files}
% \label{sec:bibtex}

% Unicode cannot be used in Bib\TeX{} entries, and some ways of typing special characters can disrupt Bib\TeX's alphabetization. The recommended way of typing special characters is shown in Table~\ref{tab:accents}.

% Please ensure that Bib\TeX{} records contain DOIs or URLs when possible, and for all the ACL materials that you reference.
% Use the \verb|doi| field for DOIs and the \verb|url| field for URLs.
% If a Bib\TeX{} entry has a URL or DOI field, the paper title in the references section will appear as a hyperlink to the paper, using the hyperref \LaTeX{} package.

\section*{Acknowledgments}

This publication has emanated from research supported in part by a
grant from Science Foundation Ireland under Grant number
SFI/12/RC/2289\_P2 \{Insight\} and a grant from Fidelity Investments. For the purpose of Open Access, the authors have
applied a CC BY public copyright license to any Author Accepted
Manuscript version arising from this submission.

\bibliography{custom}

\appendix

% \section{Example Appendix}
% \label{sec:appendix}

% This is an appendix.

\end{document}